\documentclass[12pt]{iopart}
\usepackage{iopams}
\usepackage{graphics}
\usepackage[next]{inputenc}
\usepackage[dvips]{epsfig}
\def\be{\begin{equation}}
\def\ee{\end{equation}}

\def\bi{\begin{itemize}}
\def\ei{\end{itemize}}
\def\bn{\begin{enumerate}}
\def\en{\end{enumerate}}
\def\bea{\begin{eqnarray}}
\def\eea{\end{eqnarray}}
\def\no{\nonumber}
\def\ba{\begin{array}}
\def\ea{\end{array}}
\def\bd{\begin{displaymath}}
\def\ed{\end{displaymath}}
\def\la{\langle}
\def\ra{\rangle}
\def\te{\theta}
\begin{document}
\title[A closed form for the interaction energy of anisotropic objects]{A closed form for the electrostatic interaction between two rod-like charged objects}

\author{M Askari$^1$ and J Abouie$^{1,2}$}
\address {$^1$Physics Department, Shahrood University of Technology,
Shahrood 36199-95161, Iran}
\address{$^2$School of physics, Institute for Research in Fundamental Sciences (IPM),
Tehran 19395-5746, Iran}

\ead{jahan@shahroodut.ac.ir}
\date{\today}

\begin{abstract}
We have calculated the electrostatic interaction between two rod-like charged objects with arbitrary orientations in three dimensions.
we obtained a closed form formula expressing the interaction energy in terms of the separation distance between the centers of the two rod-like objects, $r$, their lengths (denoted by $2l_1$ and $2l_2$), and their relative orientations (indicated by $\te$ and $\phi$).
When the objects have the same length ($2l_1=2l_2=l$), for particular values of separations, i.e for $r\leq0.8 l$, two types
of minimum are appeared in the interaction energy with respect to $\theta$.
By employing the closed form formula and introducing a scaled temperature $t$, we have also studied the thermodynamic properties of a
one dimensional system of rod-like charged objects.
For different separation distances, the dependence of the specific heat of the system to the scaled temperature has been studied. It is found that for $r<0.8 l$, the specific heat has a maximum.
\end{abstract}

\maketitle

\section{Introduction}
Electrostatic interactions between rod-like and stripe-like charged objects play a substantial role in the many
systems of condensed matter and soft condensed matter physics such as
strongly correlated materials, liquid crystals, electrolytes, polymers, and bio-molecules.
For example, superconductors and anti-ferromagnetic Mott insulators are a vast class of
strongly correlated systems which their quantum phases have been studied intensively.
Several evidences signifying the segregation of charged carriers into fluctuating stripes in cuprate superconductors have been augmented \cite{Tran95,Tran99,Waki99,Riga98}. In the ceramic samples ${\rm La}_{2-x}{\rm Sr}_x{\rm CuO}_4$ co-doped with ${\rm Nd}$ the charged stripes are seen by neutron diffraction as a static lattice modulation coherent with the low-temperature tetragonal modulation \cite{Tran95,Tran99}.
Nickelates \cite{Tran95} (${\rm La}_{2-x}{\rm Sr}_x{\rm NiO}_4$) are the other good empirical realities for observing the charged stripes at low temperatures.
The relationship between correlations of charged stripes and high-$T_c$ superconductivity has been studied intensively in nickelates.
More physical insights are gained by obtaining a closed form formula for the interaction of such stripes.
Indeed, the closed form gives us many useful information on the physical properties of the system such as orderings and correlations of the charged stripes and also the thermodynamic properties of the system.

Experimental evidences of such interactions are also observed in the colloidal world. Actually in the colloidal solutions, macromolecules such as DNA molecules, ${\rm TMV}$ or ${\rm fd}$ virus, ${\rm V}_2{\rm O}_5$ ribbons, crystalline Boehmite (AlOOH) rods, etc.\cite{Frad89,Purd03,Guil02,Pell99} are intrinsically very anisotropic and have rod-like and ribbon-like shapes.
In the context of orientational phase transition, it has been elucidated that the electrostatic interaction between the polyelectrolytes is lead to a twisting effect which enhances the concentration at the isotropic to nematic (I-N) phase transition \cite{{Stro86}, Odij86}.
More findings are attained by making use of the form of the interaction energy between such objects.
Many efforts have been done to employ the analytical and numerical approaches for computing
the electrostatic interaction of two rod-like charged objects.
In a route developed work, D. Chapot et. al considered two anisotropic particles, separated by a large distance (i.e a distance larger than their typical dimension) and generalized the Derjaguin, Landau, Verwey and Overbeek ({\rm DLVO}) calculations for the anisotropic molecules \cite{Verw48}. They have obtained an expression for the interaction energy between the particles \cite{Chap03}. However, because of performing various expansions, their expression is valid in the far field limit, i.e for inter-particles distance larger than both the Debye length $(\lambda_{\rm D})$ and the typical size of the colloid $a$. Meanwhile, the surface charge of the objects is involved in their calculations which should be obtained numerically. In the other hand, using molecular dynamics and Monte carlo simulations the interplay between the electrostatic interactions and orientational correlations have been studied for a 1D array of charged rods \cite{Fazli05}.

Despite of all simulations and analytical approaches, the lake of a closed form formula for the interaction energy of such
anisotropic particles is felt.
In this paper we have obtained a \textit{closed form} formula for the interaction energy
of two rod-like charged objects with lengths $2l_1$ and $2l_2$ and arbitrary orientations in three dimension, separated by $r$.
By using the closed form formula we have computed the thermodynamic functions of a one dimensional array of charged rods in which each rod interacts only with two nearest neighbors. Although, the electrostatic interaction is long range and there are no practical realization to vindicate our assumption, one can get more insight about our finding closed form formula and its direct effects on the thermodynamic behaviors of a system.
Finally the relationship between the behavior of the specific heat and the electrostatic interaction energy is investigated.
\section{Computational Details}
\par
\begin{figure}[h]
\centerline{\includegraphics[width=10cm,angle=0]{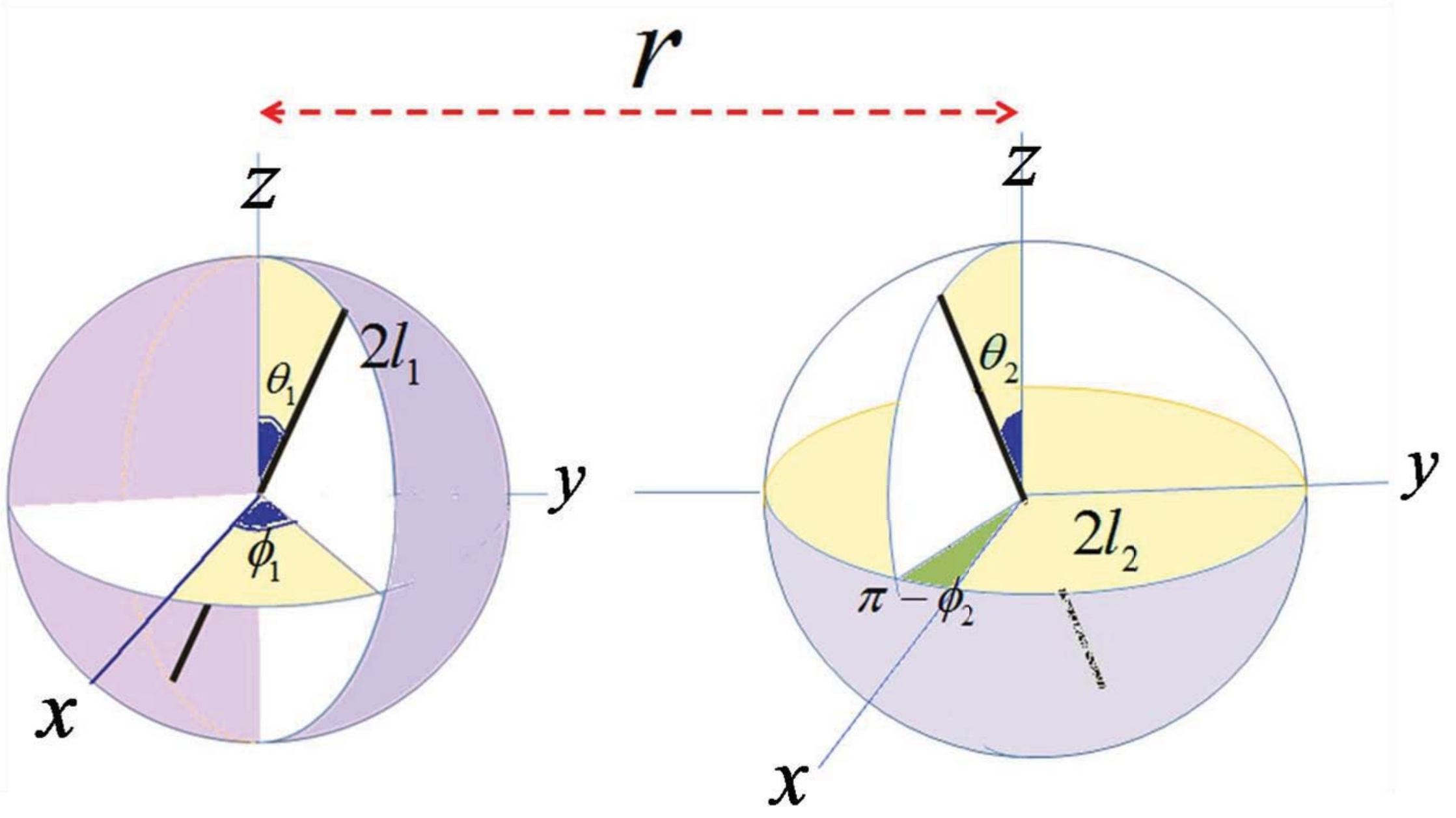}}
\caption{Schematic representation of the orientations of two rods, separated by a distance $r$. The orientations of each rod are indicated by two types of angle. $\te_{1(2)}$ are the angles of the rods with $z$ axis (co-latitude angles) and $\phi_{1(2)}$ are the angles of $l_{1(2)}\sin\te_{1(2)}$ with $x$ axis (azimuthal angles).} \label{rod}
\end{figure}

Let us consider a system of two inflexible rod-like charged objects with lengths $2l_1$ and $2l_2$, which are separated by a distance $r$.
The interaction energy between two such uniformly charged objects is given by the following formula \cite{Jackson}
\be
\mathbf{E}_{int.}=\frac{\lambda_1\lambda_2}{4\pi\varepsilon}\int_{-l_1}^{l_1}\int_{-l_2}^{l_2}\frac{e^{-\kappa_{\rm D}|{\bf r}_2-{\bf r}_1|}dl_1dl_2}{|{\bf r}_2-{\bf r}_1|}
\label{Formula1}
\ee
where ${\bf r}_1$ and ${\bf r}_2$ indicate the position vectors of two charge elements $dq_1=\lambda_1dl_1$ and $dq_2=\lambda_2dl_2$, respectively.
The inverse Debye screening length, $\kappa_{D}$, does characterize the decay rate of the interaction energy in the far field limit.
This provides a way to find the screening factor from interaction force measurement.
To calculate the interaction energy of two such rods it is convenient to work in the spherical coordinates (Fig.(\ref{rod})).
Without lose of generality, we can set up the coordinate system in which the first rod is centered on the origin and aligned along the $z$ axis and the second rod makes the angle $\theta$ with $z$ axis. Meanwhile the projection of the second rod on the $xy$ plane makes the angle $\phi$ with $x$ axis. $\theta$ and $\phi$ are related to the old coordinates by
$\cos\theta=\cos\theta_1\cos\theta_2+\sin\theta_1\sin\theta_2\cos(\phi_2-\phi_1)$ and $\phi=\phi_2-\phi_1$.
Indeed $\cos\te$ is $\hat{\mathbf{t}}_1\cdot\hat{\mathbf{t}}_2$ where $\hat{\mathbf{t}}_1$ and $\hat{\mathbf{t}}_2$ are two unit vectors
in the rods' directions and $\la\hat{\mathbf{t}}_1\cdot\hat{\mathbf{t}}_2\ra=\la\cos\te\ra$ shows the correlation between the two rods' orientations.
By introducing the polar coordinates for $dl_1$ and $dl_2$, M. Fixman and J. Skolnick calculated the screened coulomb interaction of two rigid rods
of length $l$, where $l$ assumed to be much larger than the Debye screening length \cite{Fixm78}. Except for a proportionality constant, they put forward the result obtained previously by Onsager \cite{Onsa49}. In the limit of large Debye screening lengths, i.e $\kappa_{\rm D}\rightarrow 0$,
the screening effects are neglected and the exponential term in the interaction energy Eq.(\ref{Formula1}) is close to one.
In this case, by defining the dimensionless interaction energy, $\mathbf{w}$,  we can write the interaction energy as following:
\begin{equation}
\nonumber\mathbf{E}_{int.}=\frac{q_1q_2 r}{4\pi\varepsilon (2l_1)(2l_2)}\mathbf{w},
\end{equation}
where,
\begin{equation}
\fl~~~~~~~~~~\mathbf{w}=\int_0^{\ell_1}\int_0^{\ell_2}\sum_{i=1,2 ; \alpha=\pm}
\frac{dz_1dz_2}{\sqrt{z_2^2~\mathrm{J}^2+\left[1-(-1)^iz_2\mathrm{H}\right]^2+[z_2\cos\te+\alpha(-1)^iz_1]^2}},~~~~~\label{IE}
\end{equation}
Here we have written four kinds of integral in a compact form, by introducing the coefficients $\alpha$'s, the indices $i$'s and
dimensionless functions: $\mathrm{H}=\sin\te\sin\phi$,
$\mathrm{J}=\sin\te\cos\phi$, $\ell_{1(2)}=\frac{l_{1(2)}}{r}$ and $dz_{1(2)}=\frac{dq_{1(2)}}{r\lambda_{1(2)}}$.
Actually we have divided each rod to two equal halves.
Then we have considered the interactions between each half a rod with the two halves of the other rod.
Using some transformations, we calculated the integrals of Eq.(\ref{IE}) analytically to attain to the closed form of the interaction energy. As an illustration let us describe the solution of the integral which is indicated by $\alpha=+$ and $i=1$. By integrating out $z_1$ the remaining expression is given by
\begin{eqnarray}
\no&&\int_0^{\ell_2}\bigg(\ln\left[z_2\cos\theta+\sqrt{1+z_2^2+2z_2\mathrm{H}}\right]\\
&&-\ln\bigg[z_2\cos\te
-\ell_1+\sqrt{1+z_2^2+\ell_1^2+2z_2\mathrm{H}-2\ell_1 z_2\cos\theta}\bigg]\bigg)dz_2,\label{IE-1}
\end{eqnarray}
The first logarithm in (\ref{IE-1}) is canceled by the same term with opposite sign in the other integrals which have not been presented here. However the second integral in  (\ref{IE-1}) is solved by part and is given by the following form
\begin{eqnarray}
\fl\ell_2\left(1-\ln\left[\ell_2\cos\te-\ell_1+\sqrt{\ell_2^2+u^2+\ell_2
x}\right]\right)
-\int_0^{\ell_2}dz_2\frac{\frac{xz_2}{2}+u^2-\ell_1f(z_2)}{f(z_2)\left[z_2\cos\te-\ell_1+f(z_2)\right]},\label{IE-2}
\end{eqnarray}
where $f(z_2)=\sqrt{z_2^2+u^2+z_2 x}$, $u^2=1+\ell_1^2$ and $x=2(\mathrm{H}-\ell_1\cos\te)$. Since $(u^2-\frac{x^2}{4})>0$, mathematically we are allowed to perform the following transformation for calculating the integral of Eq.(\ref{IE-2})
\begin{equation}
z_2+\frac{x}{2}=\sqrt{u^2-\frac{x^2}{4}}\left(\frac{2y}{1-y^2}\right),\label{IE-3}
\end{equation}
where $y$ is an auxiliary variable. By employing the transformation (\ref{IE-3}) we can solve the integral of (\ref{IE-2}), simply (See the appendix).

By making use of the same procedure to solve the other integrals in Eq. (\ref{IE}) i.e different $\alpha$ and $i$, we obtain a closed form formula for the interaction energy in terms of the rods' lengths, their separations, and the angles $\theta$ and $\phi$, which specify the relative orientation of two rods.
\begin{eqnarray}
\no&&\fl\mathbf{E}_{int.}=\frac{q_1q_2 r}{4\pi\varepsilon (2l_1)(2l_2)}\mathbf{w},\\
\nonumber&&\fl\mathbf{w}=\ell_2\sum_{n=0}^1\ln\left[\frac{\ell_2\cos\te+\ell_1+\sqrt{\ell_1^2+\ell_2^2+1+2\ell_2\left[(-1)^n \mathrm{H}+\ell_1\cos\te\right]}}
{\ell_2\cos\te-\ell_1+\sqrt{\ell_1^2+\ell_2^2+1-2\ell_2\left[(-1)^n \mathrm{H}+\ell_1\cos\te\right]}}\right]\\
\no&&\fl+\sum_{n,m=0}^1\sum_{j,k=1}^2(-1)^{n+m}\left[\frac{\mathrm{H}}{1+(-1)^k\cos\te}+(-1)^{k+m}\ell_1\right]\times\\
\no&&\ln\left[\tanh\left[\frac{1}{2}\sinh^{-1}\left[\frac{n\ell_2+(-1)^{j+m}\mathrm{H}+(-1)^{j}\ell_1\cos\te}{\mathrm{P}_2^m}\right]\right]-(-1)^k\right]\\
\no&&\fl+\frac{2\cos\phi}{\sin\te}\sum_{n,m=0}^1\sum_{j=1}^2(-1)^{n+j}
\tan^{-1}\bigg[\frac{1}{\mathrm{J}}\bigg(\left[\mathrm{P}_1^m+(-1)^j\ell_1\sin^2\te+(-1)^{j+m}\frac{\mathrm{F}}{2}\right]\times\\
\no&&\tanh\left[\frac{1}{2}\sinh^{-1}\left[\frac{n\ell_2
+(-1)^{j+m}\mathrm{H}-(-1)^j\ell_1\cos\te}{\mathrm{P}_1^m}\right]\right]+\mathrm{P}_1^m\cos\te\bigg)\bigg]\\
&&\no\fl+\frac{\sin\phi}{\sin\te}
\sum_{m,n=0}^1\ln\frac{\left(A_{m,n}
\tanh\left[\frac{1}{2}\sinh^{-1}[\frac{m\ell_2+(-1)^{n+m}\mathrm{H}-(-1)^{n}\ell_1\cos\te}{\mathrm{P}_1^m}]\right]+
\mathrm{P}_1^m\cos\te\right)^2+\mathrm{J}^2}{\left(B_{m,n}
\tanh\left[\frac{1}{2}\sinh^{-1}[\frac{m\ell_2-(-1)^{n+m}\mathrm{H}-(-1)^{n}\ell_1\cos\te}{\mathrm{P}_2^m}]\right]+
\mathrm{P}_2^m\cos\te\right)^2+\mathrm{J}^2},\\&&\label{C-Form}
\end{eqnarray}
where
\bea
\no&& A_{m,n}=\mathrm{P}_1^m+(-1)^n\ell_1\sin^2\te+(-1)^{n+m}\frac{\mathrm{F}}{2},\\
\no&& B_{m,n}=\mathrm{P}_2^m+(-1)^n\ell_1\sin^2\te-(-1)^{n+m}\frac{\mathrm{F}}{2},\\
\no&& \mathrm{P}_1^m=(1-m)\mathrm{P}^++m\mathrm{P}^-,\\
\no&& \mathrm{P}_2^m=(1-m)\mathrm{P}^-+m\mathrm{P}^+,\\
\no&& \mathrm{P}^{\pm}=\sqrt{\mathrm{G}\pm\ell_1\mathrm{F}},\\
\no&& \mathrm{G}=1-\sin^2\te(\sin^2\phi-\ell_1^2),\\
\no&& \mathrm{F}=\sin[2\te]\sin\phi.
\eea
As it is explicitly seen from Eq. (\ref{C-Form}) the closed form formula depends on the variables $r$, $\te$ and $\phi$.
This formula is utilized as the starting point for the most computations in the various subjects of
condensed matter and soft condensed matter physics such as the field of strongly correlated materials, liquid crystals, polymers and biophysics. Employing the closed form formula (\ref{C-Form}) we can investigate the thermodynamic properties, the order parameters
and the different phases of a system, more precisely. It should be noticed that the above expression for the interaction energy of two rod-like particles is realistic for separations larger than the particles size, i.e the separations larger than the diameter or lateral extension of the particles.

\section{Results and Discussions}
\begin{figure}[h]
\centerline{\includegraphics[width=6.cm,angle=0]{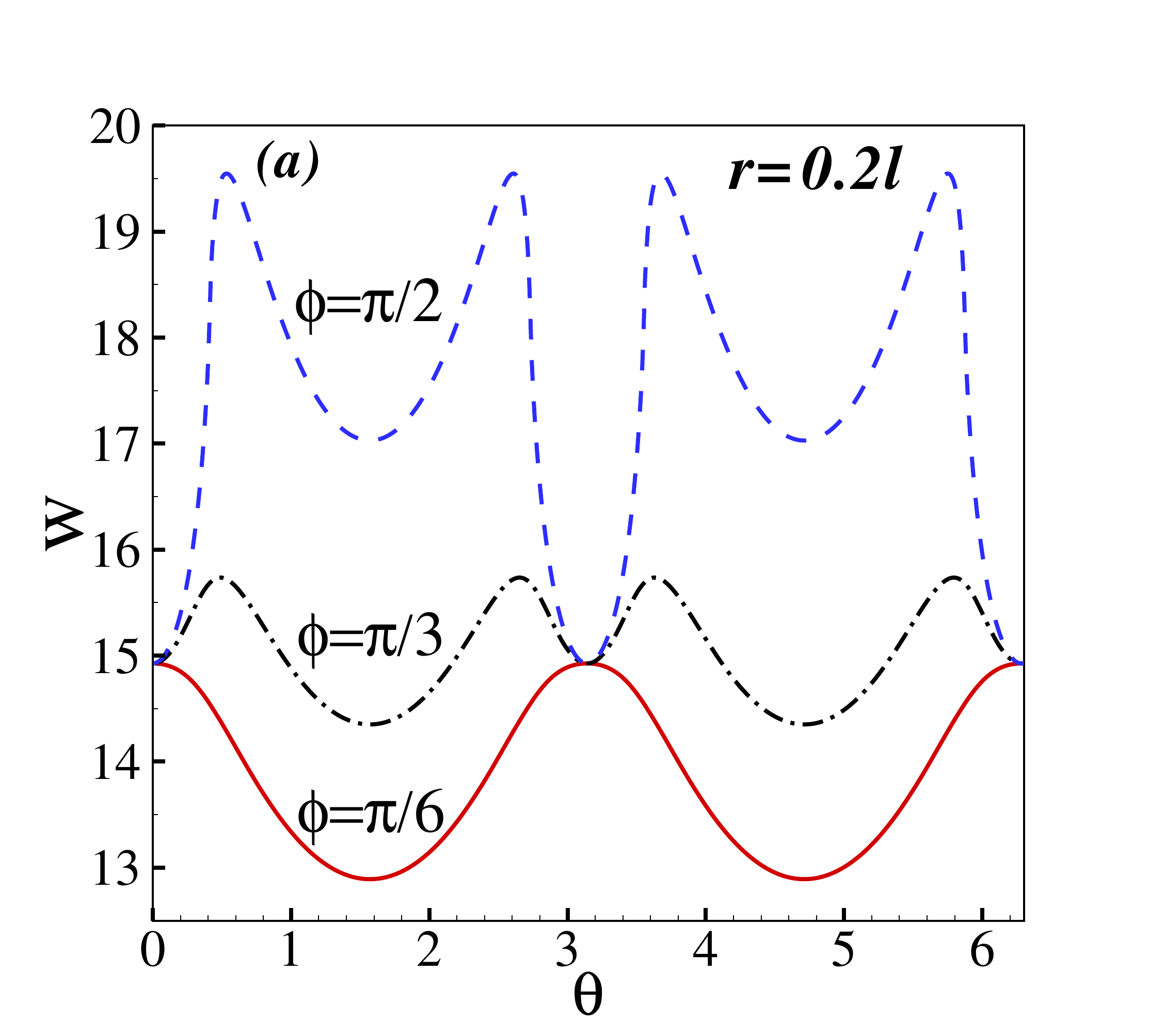}\includegraphics[width=6.cm,angle=0]{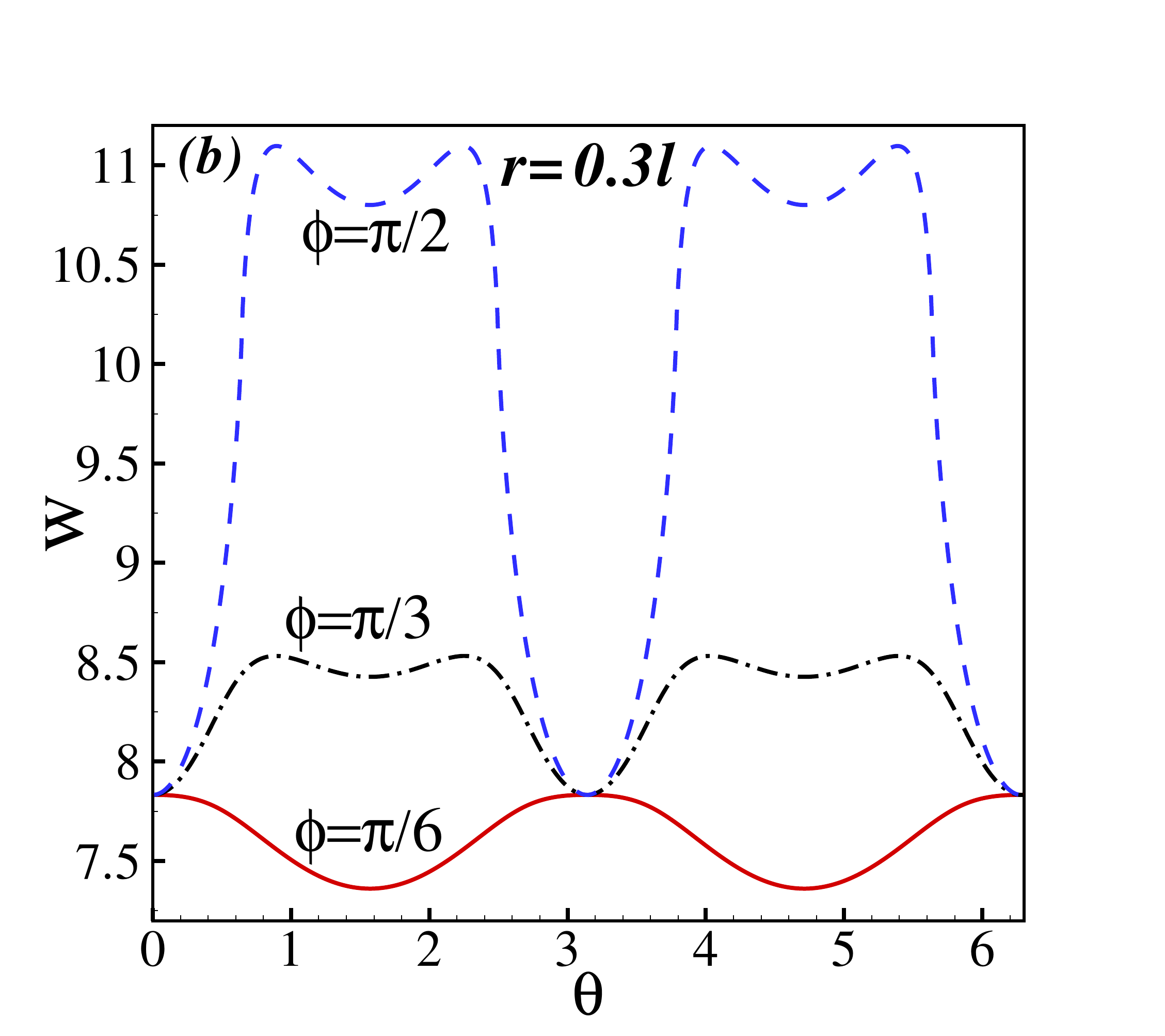}}
\centerline{\includegraphics[width=6.cm,angle=0]{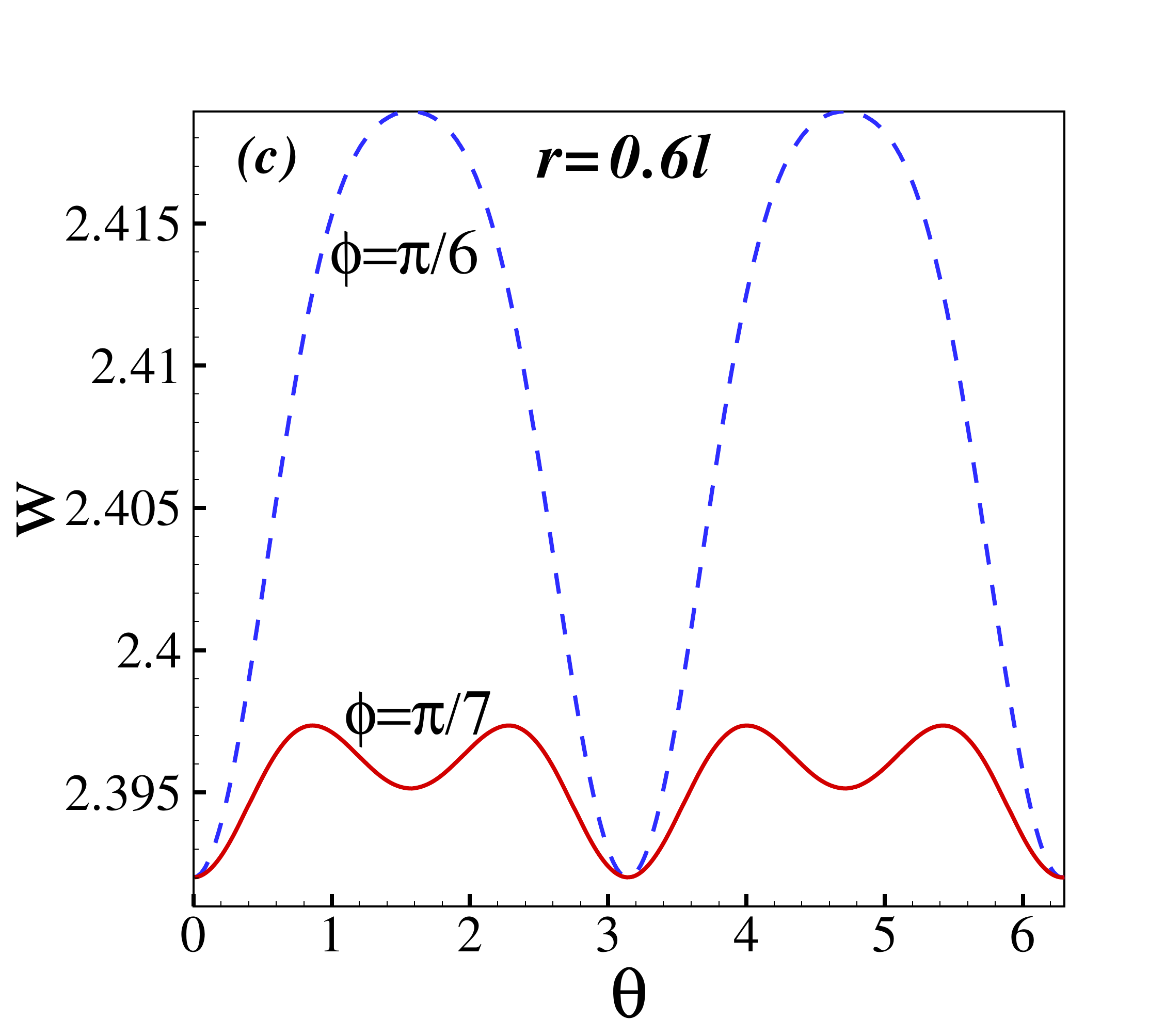}\includegraphics[width=6.cm,angle=0]{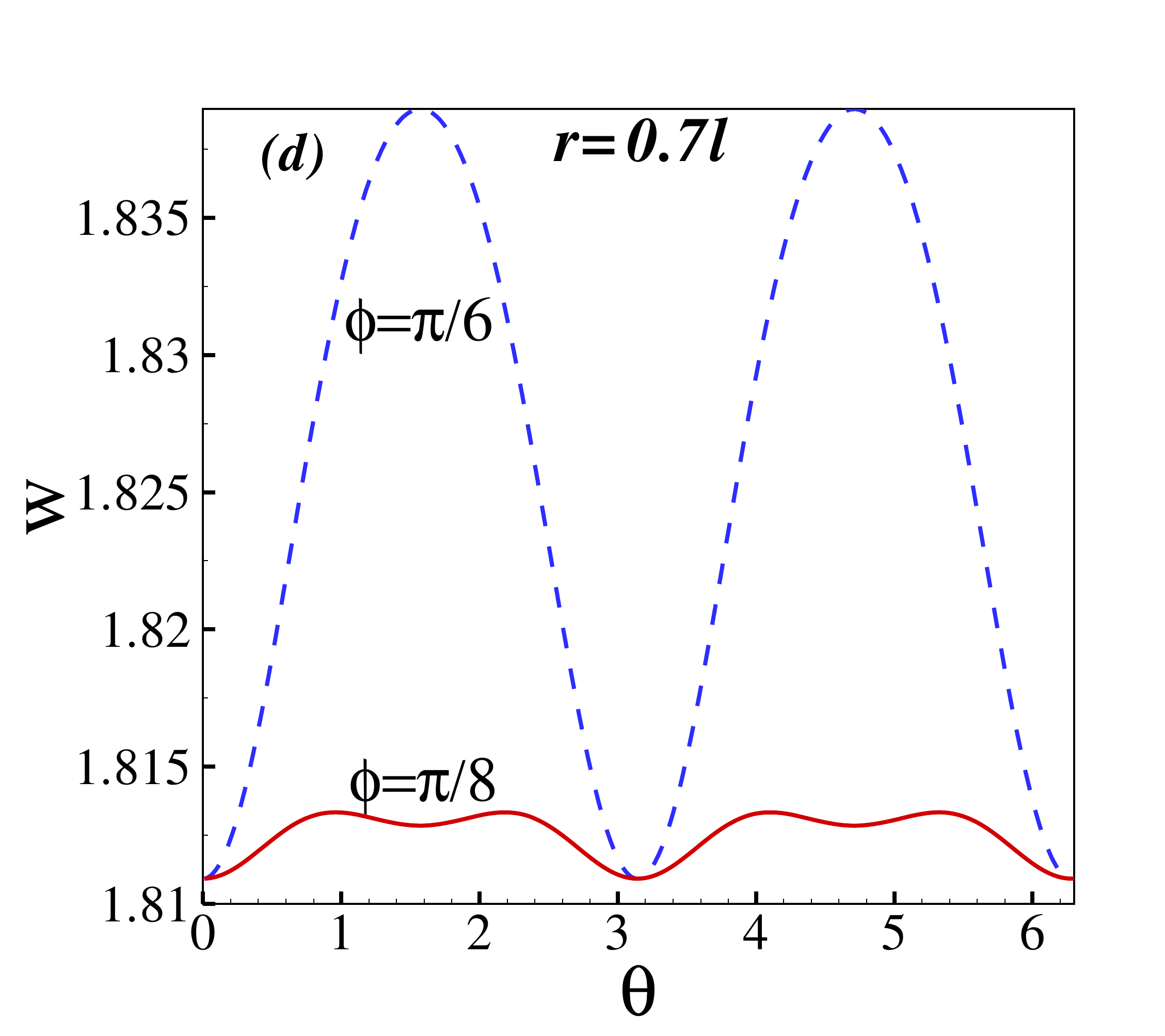}}
\caption{The interaction energy between two rods with equal lengths $2l_1=2l_2=l=8$ versus $\te$ and different values of $\phi$ and $r$.
}\label{EIE}
\end{figure}

Let us scrutinize the different rods' orientations with respect to their interaction energy.
In Fig.(\ref{EIE}) the scaled interaction energy have been plotted versus $\te$ and different values of $\phi$ and $r$.
Let us focus on the solid line curve corresponding to $\phi=\frac{\pi}{6}$ and $r=0.2 l$ in Fig.(\ref{EIE}-$a$).
Considering $n$ as an integer number, it is clearly seen that the interaction energy for the case of $\te=n\pi$ is stronger than for the case of
$\te\neq n\pi$ and for the case of $\te=\frac{2n+1}{2}\pi$ is weaker than the case of $\te\neq \frac{2n+1}{2}\pi$.
Moreover, the configurations indicated by  $(\te=n\pi,\phi=\frac{\pi}{6}, r=0.2 l)$ and $(\te=\frac{2n+1}{2}\pi,\phi=\frac{\pi}{6}, r=0.2 l)$
are the maximum and the minimum energy configurations, respectively.
By increasing $\phi$ the interaction energy is heightened, except for the directions of $\te=n\pi$.
Indeed the situation denoted by $\te=n\pi$ corresponds to the parallel rods and does not depend on $\phi$.
Thus no variation is seen on the energy value of the two rods and $\mathrm{w}(\te=n\pi,\phi, r=0.2 l)\simeq15$.
For the orientations which are indicated by $\te\neq n\pi$, an enhancement of $\phi$
makes one side of the tilted rod closer to the fixed rod and the interaction energy is increased.
The appearing of two minimums, viz. the local and the global minimums
in the interaction energy are noticeable (see Fig.\ref{EIE}-$a$-dashed line).
Therefore by increasing $\phi$ two minimums are emerged on the interaction energy at $\te=\pi$ (when two rods are parallel)
and $\te=\frac{\pi}{2}$ (when tow rods have been lied on two individual perpendicular planes).
To see the configurations with the local or the global minimum energy,
we have divided the whole region of $\phi\in[0,\pi]$ to the five intervals (see the table (\ref{tab1})).
\begin{table}[h]
\caption{The minimum and the maximum energy orientations for the rods which have been separated by $r=0.2 l$. G-Min: global minimum and L-Min: local minimum. The G-Min value is smaller than the L-Min one.}
\label{tab1}
\footnotesize\rm
\begin{tabular*}{\textwidth}{@{}l*{15}{@{\extracolsep{0pt plus12pt}}l}}
\br
$\phi$&$[0,\frac{\pi}{5}]$&$(\frac{\pi}{5},\frac{\pi}{2})$&$[\frac{\pi}{2},\frac{5\pi}{8})$&$[\frac{5\pi}{8},\frac{3\pi}{4}]$&$(\frac{3\pi}{4},\pi]$\\
\mr
$\te=n\pi$&Max.&L-Min. &G-Min.&L-Min.&Max.\\
$\te=\frac{2n+1}{2}\pi$&Min.&G-Min. &L-Min.&G-Min.&Min.\\
\br
\end{tabular*}
\end{table}

Let us compare the interaction energy between two parallel and two perpendicular rods which are separated by the distance $r=0.2 l$.
As it is obviously seen from the table (\ref{tab1}) at the interval $\phi\in[\frac{\pi}{2}, \frac{5\pi}{8})$ the parallel orientation is the global minimum energy configuration, i.e the rods prefer to be parallel, however in the other intervals of $\phi$ the global minimum energy is gained when two such rods are perpendicular.
These intervals are not universal and depend on the distance of the separations.
To get more insightful information we have also plotted in Fig.(\ref{EIE}-$b$, $c$ and $d$)
the scaled energy $\mathrm{w}$ versus $\te$ for different separation distances $r$.
For $r>0.7 l$, the orientations indicated by $(\te=n\pi, \phi)$ and $(\te=\frac{2n+1}{2}\pi,\phi)$
are always the minimum and the maximum energy configurations, respectively.
Thus for $r>0.7 l$ two rods prefer to be parallel.
By decreasing the distance of the separation from $0.7 l$ the relative orientations corresponding to the maximum and the minimum energies are changed.
\begin{figure}[h]
\centerline{\includegraphics[width=7.cm,angle=0]{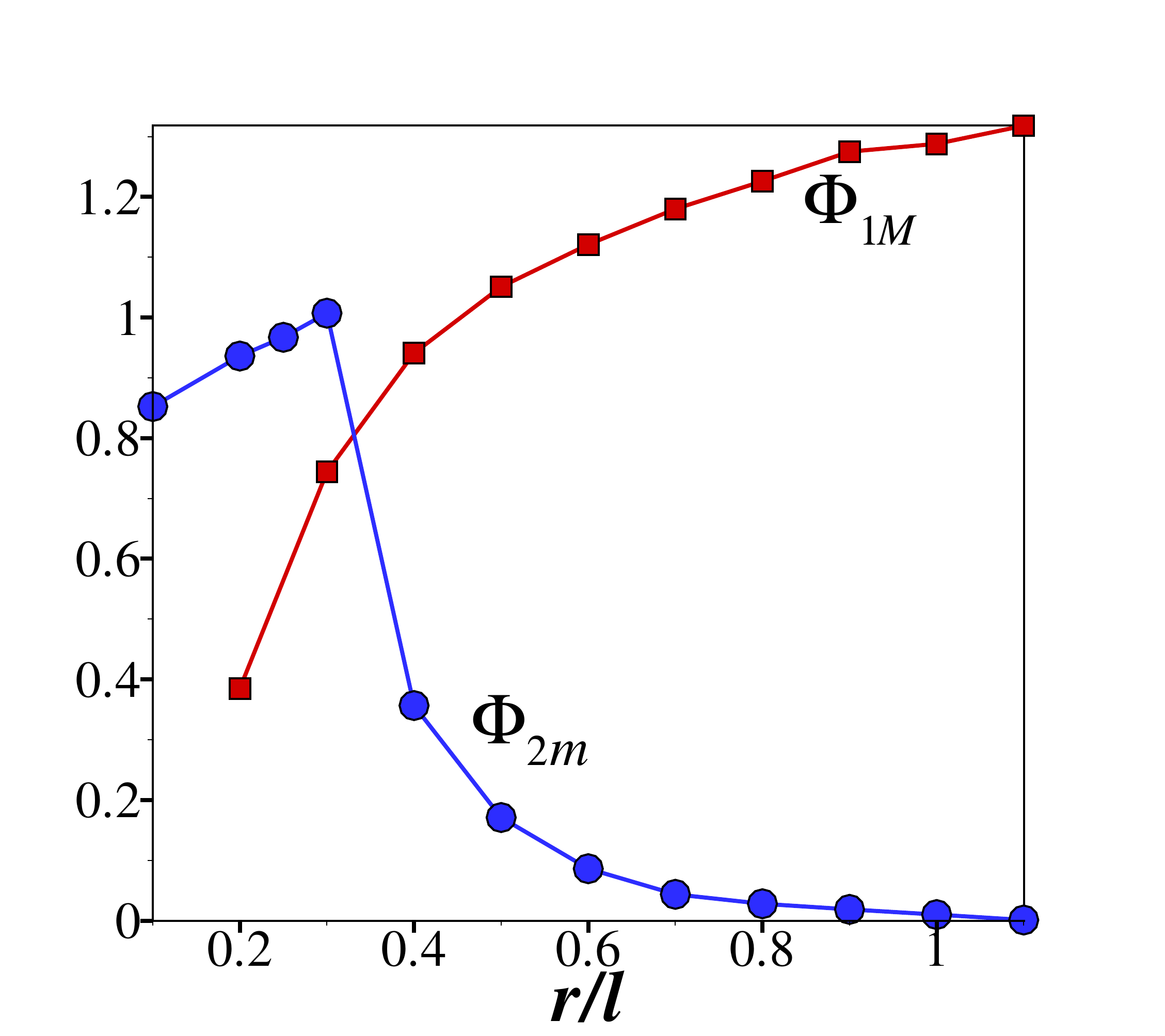}}
\caption{$\Phi_{1M}$: The interval of $\phi$ where the orientations $(\te=n\pi, \Phi_{1M})$ are the global minimum and $\Phi_{2m}$: The interval where two minimums are appeared in the interacting energy.} \label{Phi}
\end{figure}

To discover the minimum energy orientations it is more beneficial to introduce an interval, say $\Phi_{1M}$ in which the orientations indicated by $\te=n\pi$ (parallel orientations) are the global minimums.
We have plotted the function $\Phi_{1M}$ versus $r$ in Fig.(\ref{Phi}).
As it is clearly seen $\Phi_{1M}$ is increased by increasing $r$ and saturated approximately at $r\sim0.8 l$.
It is also useful to define another interval, $\Phi_{2m}$, where the state of the system is characterized by two
minimums in the interaction energy. The corresponding relative orientations are
indicated by $(\te=n\pi, \Phi_{2m})$ and $(\te=\frac{2n+1}{2}\pi, \Phi_{2m})$.
As it is also seen from the Fig.(\ref{Phi}) by decreasing the separated distance $r$,
the intervals of $\Phi_{2m}$ are increased and reached to a maximum value at $r\sim0.3 l$.
It is remarkable that the interval $\Phi_{2m}$ has considerable values only for $r\leq0.8 l$.
This characteristic behavior gives an insightful intuition when we discuss the response function of a 1D array of the interacting rods.
\begin{figure}[h]
\centerline{\includegraphics[width=8cm,angle=0]{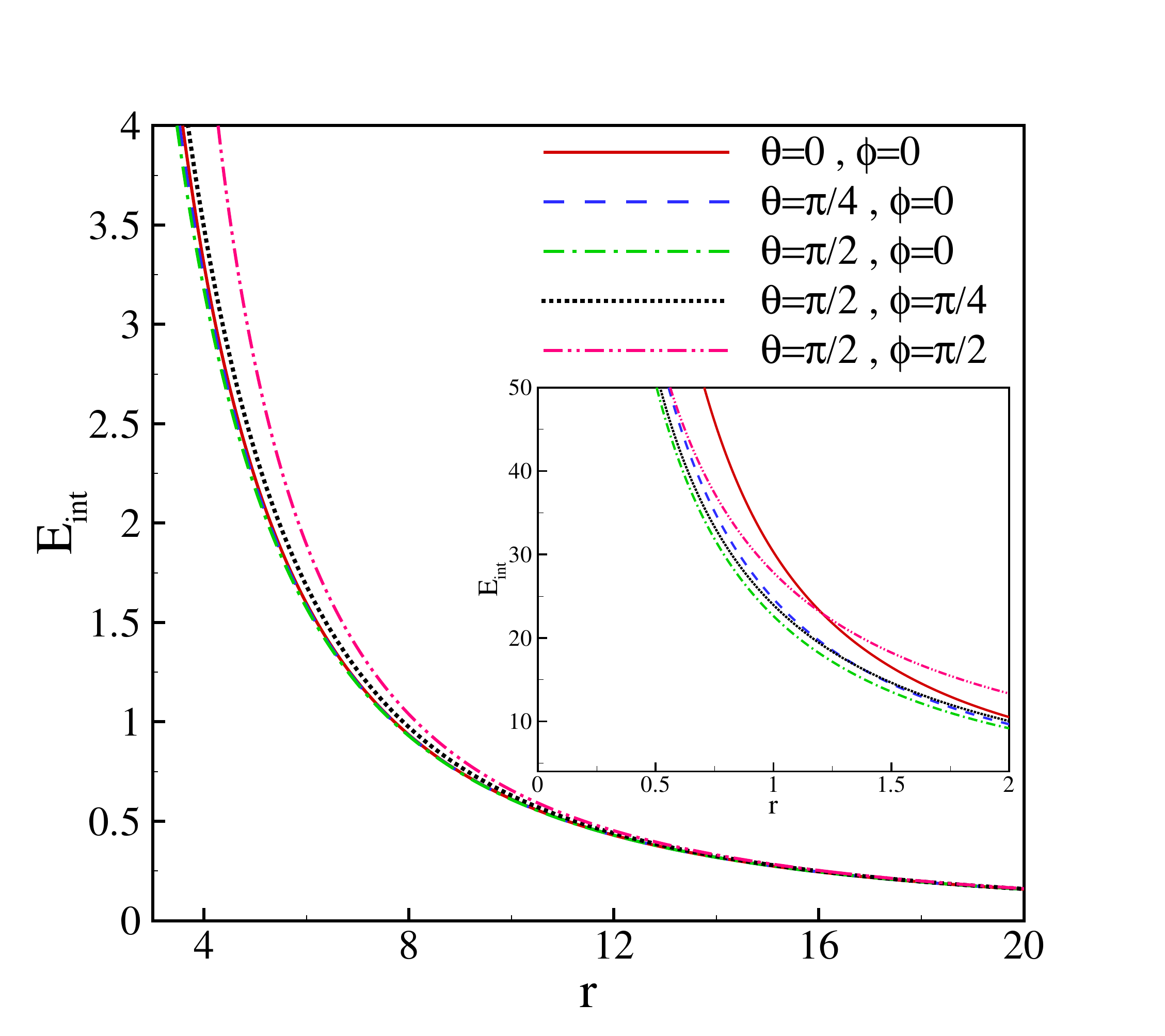}}
\caption{The interaction energy between the rods $(l=8)$ versus $r$ and different relative orientations.} \label{IE-Fig}
\end{figure}

Fig.(\ref{IE-Fig}) shows the interaction energy of two charged rods as a function
of their separations $r$, and different relative orientations.
The interaction energy decreases with increasing the distance of separation between the rods.
It is seen that all curves coincide at large $r$ values, indicating that at large values of $r$, the rods can be treated as point charges,
which is in well agreement with other works\cite{Chap03}.
\begin{figure}[h]
\centerline{\includegraphics[width=7.cm,angle=0]{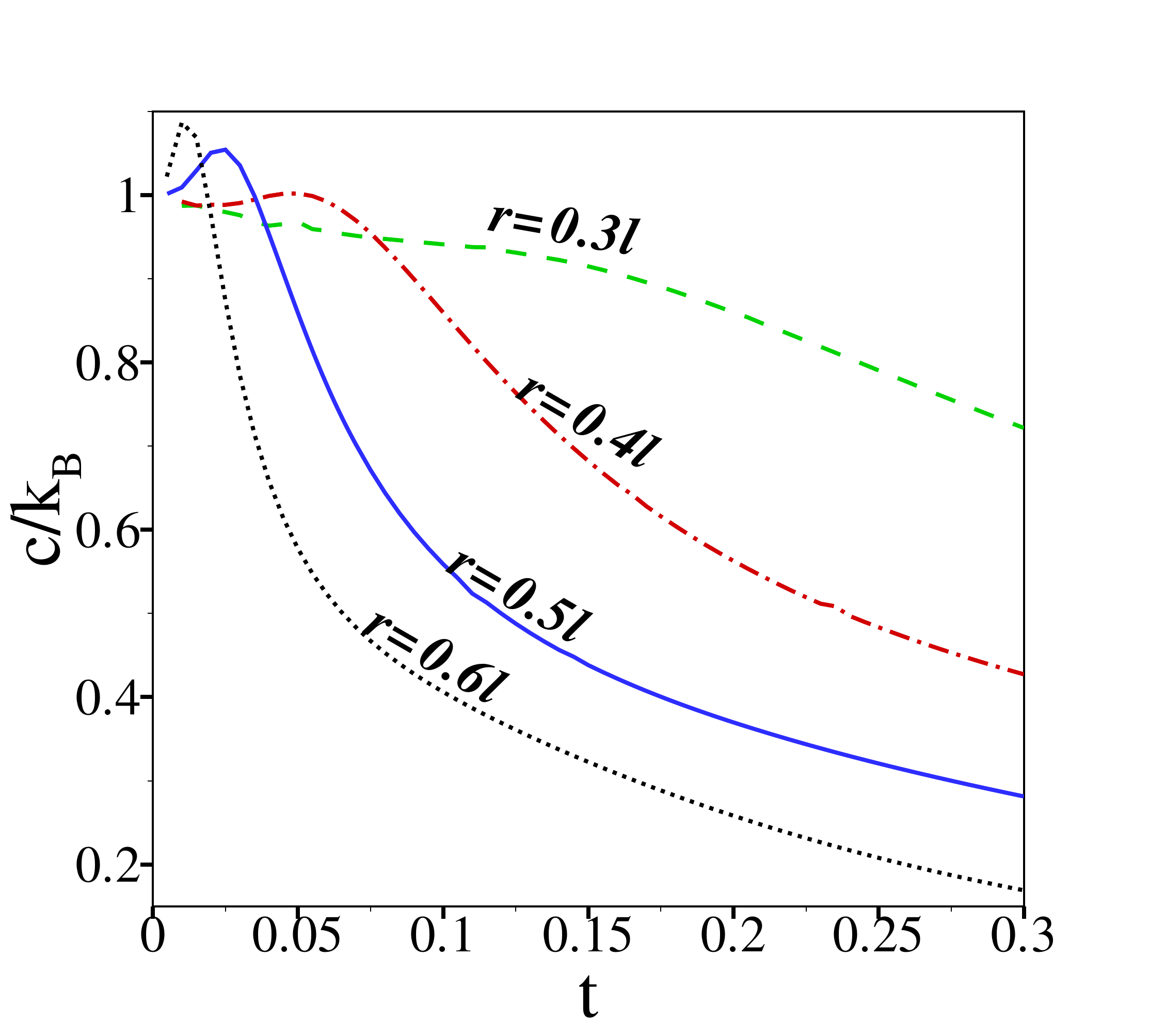}\includegraphics[width=7.cm,angle=0]{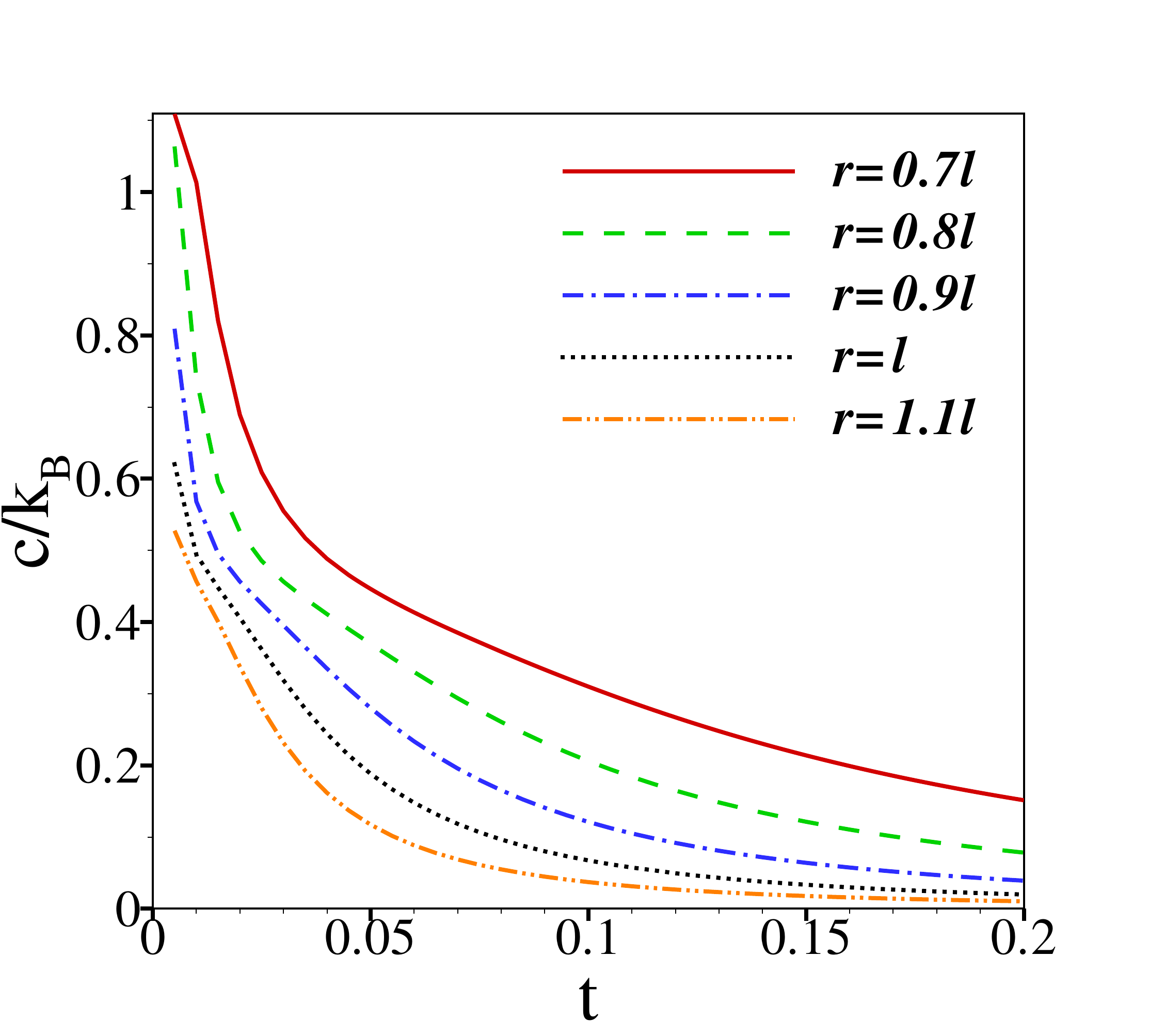}}
\caption{The specific heat for a chain of N.N interacting rods ($l=8$). Right: for $r< 0.7 l$ where two types of minimum are appeared in the interacting energy plot. Left: for $r\geq0.7 l$} \label{spe-heat}
\end{figure}

To see the advantage of the closed form formula, let us consider a 1D array of interacting rods which are separated by a lattice constant $r$.
Typically this model is used to discuss the properties of the 1D liquid crystals. We have considered the situation
in which each rod interacts only with its nearest neighbors (N.N). Indeed, the electrostatic interaction is long-range and there are no practical realization to vindicate our assumption. However we bring this example, to show the direct influence of the closed form formula (\ref{C-Form}) on the thermodynamic behaviors of the system.

We have defined a re-scaled temperature, $t=\frac{k_{\rm B}T}{\frac{q_1q_2}{4\pi\varepsilon r\ell^2}}$, where $T$ is the temperature, $k_{\rm B}$ is the Boltzmann constant, and $\ell=\frac{l}{r}$ is a dimensionless length ($l=2l_1=2l_2$).
The re-scaled temperature can be expressed in terms of Bjerrum length $\ell_{\rm B}=e^2/4\pi\varepsilon k_{\rm B}T$ (For water $\ell_{\rm B}=7{\AA}$ at room temperature). It can be seen that $t=\frac{r}{\ell_{\rm B}}\left(\frac{\ell}{n}\right)^2$, where $n$ is the number of fundamental charges ($e$) on the rods. Fig.(\ref{spe-heat}) shows the change in the specific heat per particle as a function of the re-scaled temperature, for a
1D array of N.N. interacting rods.

Let us construe the behavior of the specific heat with respect to the interaction energy.
As it is obviously seen from Fig. (\ref{spe-heat}), For distances $r<0.7 l$ there is a peak on the specific heat at low temperatures. However, this is not the case of $r>0.7 l$ and the peak disappears at low temperatures.
Indeed this peak is seen when there exist two types of minimum in the interaction energy.
Moreover, the width of the peak depends inversely to the value of $\Phi_{2m}$,
i.e at large values of $\Phi_{2m}$, such a peak is very broaden and becomes narrower by decreasing the interval $\Phi_{2m}$.

For more sense, it is worth to discuss the peak values and peak locations in the actual units.
As it is observed from the specific heat data, for the separation $r=0.4 l~ (\ell=2.5)$
the peak value and peak location are $c\sim 1.02~k_{\rm B}$ and $t\sim0.06$, respectively.
For a water solution, and highly charged particles $\frac{\ell}{n}\sim0.5$ and the peak location corresponds to the temperature $T\sim 10~ K$.
It means that at $T\sim 10~ K$ the specific heat for a 1D array of N.N rods has its maximum value. In the case of $\frac{\ell}{n}\sim1$ the peak occurs at $T=40~ K$.

It is remarkable again that for a realistic model we should consider the interaction between more neighbors. Discussing on the thermodynamic behaviors of such models by using our closed form formula is our future work.
\section*{Appendix}
Using the transformation (\ref{IE-3}), we can find the following expression for the integral of Eq.(\ref{IE-2}):
\begin{eqnarray}
\nonumber&&\fl\mathbf{w}=\ell_2\left(1-\ln\left[\ell_2\cos\te-\ell_1+
\sqrt{\ell_1^2+\ell_2^2+1-2\ell_2\left[\mathrm{H}+\ell_1\cos\te\right]}\right]\right)\\
\no&&\fl+\sum_{n=0}^1\left[\frac{\mathrm{H}}{1+(-1)^n\cos\te}-(-1)^n\ell_1\right]
\ln\left[\frac{\tanh\left[\frac{1}{2}\sinh^{-1}\left[\frac{\ell_2+\mathrm{H}-\ell_1\cos\te}{\mathrm{P}^+}\right]\right]-(-1)^n}
{\tanh\left[\frac{1}{2}\sinh^{-1}\left[\frac{\mathrm{H}-\ell_1\cos\te}{\mathrm{P}^+}\right]\right]-(-1)^n}\right]\\
\no&&\fl+\frac{2\cos\phi}{\sin\te}\sum_{n=0}^1(-1)^n\tan^{-1}\bigg[\frac{1}{\mathrm{J}}\bigg(A
\tanh\left[\frac{1}{2}\sinh^{-1}\left[\frac{n\ell_2
+\mathrm{H}-\ell_1\cos\te}{\mathrm{P}^+}\right]\right]+\mathrm{P}^+\cos\te\bigg)\bigg]\\
&&\fl+\frac{\sin\phi}{\sin\te}\ln\left[\frac{\left(A
\tanh\left[\frac{1}{2}\sinh^{-1}[\frac{\mathrm{H}-\ell_1\cos\te}{\mathrm{P}^+}]\right]+
\mathrm{P}^+\cos\te\right)^2+\mathrm{J}^2}{\left(A
\tanh\left[\frac{1}{2}\sinh^{-1}[\frac{\ell_2+\mathrm{H}-\ell_1\cos\te}{\mathrm{P}^+}]\right]+
\mathrm{P}^+\cos\te\right)^2+\mathrm{J}^2}\right]\label{C1-Form}
\end{eqnarray}
where $A=\mathrm{P}^++\ell_1\sin^2\te+\frac{\mathrm{F}}{2}$.

\ack
It is our pleasure to acknowledge H. Fazli and A. Naji for their useful comments and discussions.
We are also indebted to S. Hessami Pilehrood for reading carefully the manuscript.
We thank Mohammad Ebrahim Ghazi for introducing some fruitful references.
MA also acknowledges
S. Reza Mousawi for his helpful comments on the mathematical calculations.


\end{document}